\documentclass[sigconf]{acmart}

\renewcommand\footnotetextcopyrightpermission[1]{} 
\AtBeginDocument{%
  }

\usepackage[ruled,vlined,linesnumbered]{algorithm2e}
\usepackage{graphicx}   
\usepackage{tabularx}

\setcitestyle{numbers}

\begin{document}

\title{DAO-Agent: Zero Knowledge-Verified Incentives for Decentralized Multi-Agent Coordination}

\author{Yihan Xia}
\email{xiayihan2023@email.szu.edu.cn}
\affiliation{%
  \institution{Shenzhen University}
  \city{Shenzhen}
  \country{China}
}

\author{Taotao Wang}
\authornote{Corresponding author.}
\email{ttwang@szu.edu.cn}
\affiliation{%
  \institution{Shenzhen University}
  \city{Shenzhen}
  \country{China}
}

\author{Wenxin Xu}
\email{xuwenxin2022@email.szu.edu.cn}
\affiliation{%
  \institution{Shenzhen University}
  \city{Shenzhen}
  \country{China}
}

\author{Shengli Zhang}
\email{zsl@szu.edu.cn}
\affiliation{%
  \institution{Shenzhen University}
  \city{Shenzhen}
  \country{China}
}

\renewcommand{\shortauthors}{Xia et al.}

\begin{abstract}

  {Autonomous Large Language Model (LLM)-based multi-agent systems have emerged as a promising paradigm for facilitating cross-application and cross-organization collaborations. These autonomous agents often operate in trustless environments, where centralized coordination faces significant challenges, such as the inability to ensure transparent contribution measurement and equitable incentive distribution. While blockchain is frequently proposed as a decentralized coordination platform, it inherently introduces high on-chain computation costs and risks exposing sensitive execution information of the agents. Consequently, the core challenge lies in enabling auditable task execution and fair incentive distribution for autonomous LLM agents in trustless environments, while simultaneously preserving their strategic privacy and minimizing on-chain costs. To address this challenge, we propose DAO-Agent, a novel framework that integrates three key technical innovations: (1) an on-chain decentralized autonomous organization (DAO) governance mechanism for transparent coordination and immutable logging; (2) a ZKP mechanism approach that enables Shapley-based contribution measurement off-chain, and (3) a hybrid on-chain/off-chain architecture that verifies ZKP-validated contribution measurements on-chain with minimal computational overhead. We implement DAO-Agent and conduct end-to-end experiments using a crypto trading task as a case study. Experimental results demonstrate that DAO-Agent achieves up to 99.9\% reduction in verification gas costs compared to naive on-chain alternatives, with constant-time verification complexity that remains stable as coalition size increases, thereby establishing a scalable foundation for agent coordination in decentralized environments.}
\end{abstract}


\keywords{Multi-Agent Systems, Blockchain, Zero-Knowledge Proofs, Shapley Value, Decentralized Autonomous Organizations}


\maketitle

\section{Introduction}
\label{sec:introduction}

Autonomous Large Language Model (LLM)-based multi-agent systems have emerged as a promising paradigm for cross-application collaborations \cite{wang2025megaagent,jimenez2025multiagent}. However, these agents often operate in trustless environments where centralized coordination faces significant challenges: ensuring transparent contribution measurement, equitable incentive distribution, and preventing strategic manipulation by rational agents.

While blockchain offers decentralized coordination through smart contracts and Decentralized Autonomous Organizations (DAOs), it introduces prohibitive on-chain computation costs and risks exposing sensitive agent execution data \cite{chambefort2024blockchain,qin2023web3dao}. Existing solutions face fundamental limitations: permissioned blockchains sacrifice decentralization \cite{bakos2022permissioned}, oracle-based systems reintroduce centralized trust \cite{makhdoom2023trust,duley2023oracle}, and current verifiable computation approaches lack practical scalability for dynamic multi-agent scenarios \cite{huo2024scalable}.

This analysis reveals a fundamental dilemma between trust and efficiency as Multi-Agent Systems (MAS) evolve towards DAOs. Traditional blockchain architectures adhere to ``Code is Law,'' offering maximal transparency but suffering from severe computational constraints that preclude complex LLM reasoning or game-theoretic calculations (e.g., $O(n!)$ complexity). Conversely, off-chain systems possess abundant compute but lack verifiable mechanisms to ensure task quality or fair reward distribution.

To resolve this, our research goal is to establish a \textbf{``Compute Off-chain, Verify On-chain''} paradigm. Specifically, we aim to achieve:
\begin{itemize}
  \item \textbf{From Unstructured to Ordered}: Bridging the gap between natural language goals and executable logic by utilizing LLMs for dynamic task planning and DAG generation.
  \item \textbf{Fairness through Game Theory}: Ensuring equitable incentive distribution through Shapley values, which axiomatically quantify each agent's marginal contribution to collaborative outcomes.
  \item \textbf{Proof as Justice}: Replacing reliance on centralized coordinators with Zero-Knowledge Proofs (ZKP), effectively transforming ``trust in intermediaries'' into ``trust in mathematics.''
\end{itemize}

We propose \textbf{DAO-Agent}, a novel framework integrating these principles. DAO-Agent employs a hybrid architecture where agents collaborate off-chain while ensuring verifiable contribution measurement through cryptographic proofs.

Our main contributions are three key technical innovations:

\begin{enumerate}
  \item \textbf{Hybrid On-chain/Off-chain Architecture:} We design a novel architecture that separates computationally intensive Shapley value calculation from lightweight on-chain verification, achieving constant $O(1)$ verification cost regardless of agent count.

  \item \textbf{Fair Contribution Measurement:} We implement a contribution measurement system based on Shapley values from cooperative game theory. This mechanism provides axiomatically fair reward allocation, making honest participation a dominant strategy for rational agents.

  \item \textbf{Verifiable ``Proof as Justice'' Architecture:} We construct a ZKP-based verification pipeline (using recursive proof composition: STARK proofs wrapped in Groth16) that allows the blockchain to verify off-chain game-theoretic computations with $O(1)$ cost, ensuring that the incentive distribution is mathematically proven to be fair without exposing private strategies.
\end{enumerate}

We implement DAO-Agent and demonstrate its effectiveness through crypto trading experiments, showing significant cost reduction while maintaining fairness and system robustness at scale. The remainder of this paper systematically develops our solution: Section~\ref{sec:preliminaries} formalizes the system model and trustless coordination challenges that motivate our architectural design; Section~\ref{sec:framework} presents our hybrid on-chain/off-chain framework that transforms abstract design goals into concrete cryptographic protocols; and Section~\ref{sec:experiments} validates our approach through comprehensive experiments demonstrating constant-cost verification and fair incentive distribution. Note that detailed ZKP construction is discussed in Section~\ref{subsec:zkp-integration}.

\section{Related Work}
\label{sec:related}

Our work intersects three research domains: multi-agent collaboration systems, blockchain-based coordination mechanisms, and zero-knowledge proof technologies. We position DAO-Agent relative to prior work in each area and highlight our contributions.

\subsection{Multi-Agent Collaboration}
\label{subsec:multi-agent}

Recent advances in large language models have enabled sophisticated multi-agent collaboration systems capable of autonomous task decomposition and execution. MegaAgent \cite{wang2025megaagent} demonstrates that LLM-based systems can scale to hundreds of agents without predefined standard operating procedures, while production systems like Anthropic's multi-agent research platform \cite{anthropic2025research} leverage multiple Claude agents for parallel search and synthesis. Building on this foundation, HiveMind \cite{xia2025hivemind} introduces contribution-guided prompt optimization using Shapley value analysis, demonstrating principled agent evaluation in centralized environments. Surveys \cite{multiagent2025survey,cognitive2025llm} identify coordination, scalability, and autonomous decision-making as central challenges. However, existing approaches predominantly assume centralized coordination or trusted execution environments, with none addressing fair contribution measurement or equitable incentive distribution in competitive multi-agent scenarios. DAO-Agent extends this line of work by introducing decentralized coordination through blockchain governance, employing game-theoretic Shapley value computation for axiomatically fair reward allocation, and achieving privacy-preserving contribution measurement through zero-knowledge proofs.

\subsection{Blockchain-Based Coordination}
\label{subsec:blockchain}

Decentralized Autonomous Organizations (DAOs) have emerged as a novel governance structure enabled by blockchain technology \cite{qin2023web3dao,cardoso2023dao}. Recent work has begun exploring DAOs for AI coordination, though primarily focusing on human-in-the-loop governance rather than fully autonomous agent collaboration \cite{capponi2025daoai}. However, DAO governance faces persistent challenges including the whale problem and collusion issues \cite{tamai2024dao}. The Shapley value \cite{shapley1953value,qin2025shapley} provides a principled approach to fair allocation, but its exponential computational complexity renders direct on-chain computation prohibitively expensive. Recent advances in DAG-based approximations \cite{xia2025hivemind} have significantly reduced computational overhead while maintaining theoretical guarantees. DAO-Agent builds upon these techniques and further addresses the challenge of trustless verification through a hybrid architecture that separates computation from verification, achieving constant verification cost ($\sim$27k gas) independent of agent count while preserving computational privacy through zero-knowledge proofs.

\subsection{Verifiable Computation and Privacy}
\label{subsec:verifiable-computation}

Zero-knowledge proofs (ZKPs) have evolved from theoretical cryptographic primitives to practical tools for blockchain scalability and privacy \cite{zkp2025survey}. Comparative analyses \cite{hacken2023zkcomparison} distinguish zkSNARKs from zkSTARKs in terms of proof size, verification cost, and trusted setup requirements. However, existing applications primarily focus on transaction validation and state transitions \cite{ethereum2025zkp,bontekoe2024zkp,amiri2024bedrock}, with limited exploration of verifiable computation for multi-party coordination. Recent advances in multi-modal learning and semantic completion techniques \cite{tu2025global} demonstrate sophisticated approaches to cross-modal data alignment and semantic understanding, which inform our approach to ensuring data integrity across heterogeneous agent outputs. DAO-Agent employs a STARK-to-SNARK proof aggregation pipeline to apply ZKPs to a novel domain: proving correctness of Shapley value computation over committed agent outputs, achieving $O(1)$ on-chain verification cost while supporting exponential off-chain computation complexity.

\section{Problem Setup}
\label{sec:preliminaries}

This section establishes the system model, formalizes core challenges in trustless multi-agent coordination, and defines formal security properties that guide the architectural design in Section~\ref{sec:framework}.

\subsection{System and Threat Model}
\label{subsec:system-model}

\textbf{System Model.} Consider a collaborative task $T$ that involves $n$ autonomous LLM agents, denoted as $\mathcal{N} = \{a_1, \ldots, a_n\}$, each possessing specialized capabilities. Each agent $a_i$ produces an output $o_i$ during the task execution, where their outputs may be interdependent. We model this as a cooperative game $\Gamma = (\mathcal{N}, v)$, where the characteristic function $v: 2^{\mathcal{N}} \to \mathbb{R}$ maps each agent coalition $S \subseteq \mathcal{N}$ to a real value $v(S)$ quantifying the collaborative value created by that subset of agents \cite{shapley1953value}. Formally, $v(S) = f_{\text{eval}}(O_S)$ where $O_S = \{o_i : a_i \in S\}$ represents the collective outputs of coalition $S$, and $f_{\text{eval}}(\cdot)$ is a task-specific evaluation function. We instantiate this framework with a cryptocurrency trading task where agents collaboratively analyze market data, synthesize trading perspectives, and generate investment strategies. In this domain, $v(S)$ quantifies the trading performance achieved by coalition $S$ based on their collective market analysis and decision-making accuracy. The marginal contribution of agent $a_i$ to coalition $S$ is quantified by $v(S \cup \{a_i\}) - v(S)$. This formulation accommodates collaborative synergies through superadditivity:
\begin{equation}
  \label{eq:superadditivity}
  v(\mathcal{N}) > \sum_{i \in \mathcal{N}} v(\{a_i\})
\end{equation}
where $v(\mathcal{N})$ denotes the grand coalition value and $v(\{a_i\})$ represents agent $i$'s standalone contribution. This inequality formalizes the collaborative synergy property where collective performance exceeds individual capabilities \cite{peleg2007introduction,suijs1998stochastic}.

\textbf{Threat Model.} Rational, self-interested agents may seek to maximize individual rewards through three primary attack vectors: \textit{(1) Output Manipulation}, where an agent submits a modified output $o'_i \neq o_i$ after observing others' contributions to inflate its perceived value; \textit{(2) Collusion}, where a coalition of agents $K \subseteq \mathcal{N}$ coordinates to unfairly redistribute rewards in their favor; and \textit{(3) Strategic Withholding}, where an agent reduces its effort contribution $e_i$ below an expected baseline $\bar{e}$ while relying on others to compensate for the deficit.

\textbf{Trust Model.} Our model assumes a secure underlying blockchain and cryptographic primitives. Specifically: agents cannot compromise the blockchain infrastructure, as Byzantine fault tolerance ensures the probability of a successful network-level attack, $\epsilon_{\text{blockchain}}$, is a negligible function of the security parameter $\lambda$ (i.e., $\epsilon_{\text{blockchain}} = \text{negl}(\lambda)$ where $\text{negl}(\lambda)$ denotes any function that decreases faster than the inverse of any polynomial in $\lambda$) \cite{zhong2023byzantine}. All cryptographic primitives (e.g., hash functions) are assumed to be secure under standard computational assumptions. The off-chain Coordinator is trusted for \textit{liveness} (i.e., it will eventually perform its computational tasks) but not for \textit{correctness}; its computational integrity is enforced cryptographically through zero-knowledge proofs, which are verified on-chain.

\subsection{Core Challenges}
\label{subsec:challenges}

Given the threat model, we identify three fundamental challenges for trustless, fair, and scalable multi-agent collaboration:

\textbf{Challenge C1 (Fair Contribution Measurement):} How to quantify each agent's contribution when outputs are interdependent and collaborative value exhibits the superadditivity property (Eq.~\ref{eq:superadditivity})? The measurement must satisfy axiomatic fairness—equal marginal contributions yield equal rewards, zero marginal contribution yields zero reward—while accounting for interaction effects.

\textbf{Challenge C2 (Verifiable Output Authenticity):} How to ensure inputs for contribution measurement are authentic and tamper-proof? The system must prevent retroactive output manipulation with negligible failure probability, even when agents observe others' contributions before finalizing their own.

\textbf{Challenge C3 (Scalable Privacy-Preserving Verification):} How to achieve verifiable fairness without prohibitive on-chain costs or revealing strategic information? Fair allocation mechanisms like Shapley values require $O(2^n)$ coalition evaluations, economically infeasible on-chain beyond $n > 6$ agents. Revealing individual outputs may expose competitive strategies.

Existing approaches face fundamental limitations: centralized coordinators lack transparency (violating C2); permissioned blockchains sacrifice decentralization; oracle-based systems reintroduce trust assumptions; direct on-chain Shapley computation violates C3's scalability requirement.

\subsection{Design Goals and Formal Guarantees}
\label{subsec:design-goals}

We address challenges C1--C3 through three design goals, each paired with a verifiable guarantee that maps the identified risk to a concrete mitigation.

\subsubsection{Goal G1: Fair and Incentive-Compatible Contribution Measurement}
\label{subsubsec:goal-fairness}

To resolve Challenge C1, DAO-Agent must discourage strategic deviation while preserving Shapley fairness across all coalitions.

\textbf{Guarantee G1 (Incentive Compatibility and Collusion Resistance).}\label{property:incentive-compatibility}
DAO-Agent makes honest participation weakly dominant for each agent and suppresses coordinated manipulation as the population grows:
\begin{enumerate}
  \item $\mathbb{E}[\text{reward}_i(\text{honest})] \geq \mathbb{E}[\text{reward}_i(\sigma'_i)]$ for every unilateral deviation $\sigma'_i$.
  \item Coalition gain satisfies $\text{gain}_K \leq \tfrac{|K|}{n} \cdot v(\mathcal{N}) + O\!\left(\tfrac{|K|^2}{n}\right)$ for any colluding subset $K \subseteq \mathcal{N}$.
\end{enumerate}
Here, $\text{reward}_i(\cdot)$ denotes the expected on-chain payout to agent $a_i$ under the strategy profile supplied as an argument; $\sigma'_i$ denotes any unilateral deviation from the honest policy; $K$ labels a colluding coalition distinct from the generic coalition symbol $S$; and $\text{gain}_K$ captures the aggregate reward uplift that coalition $K$ can realize relative to honest execution.

Shapley allocation (Section~\ref{subsec:shapley}) enforces axiomatic fairness, eliminating profitable unilateral deviations when computation is faithful. Guarantee~G2 (\ref{property:commitment-binding}) makes tampering detectable with overwhelming probability, so deviations yield no reward. The coalition bound follows from the Shapley efficiency axiom: total surplus remains $v(\mathcal{N})$, turning collusion into a zero-sum redistribution that vanishes as $n$ increases.

  {
    \begin{theorem}[Incentive Compatibility]
      Under the Shapley allocation $\{\phi_i\}$ with cryptographic commitment binding (G2), honest participation is a weakly dominant strategy for each agent $a_i \in \mathcal{N}$.
    \end{theorem}

    \begin{proof}[Proof Sketch]
      By the Shapley efficiency axiom, $\sum_{i \in \mathcal{N}} \phi_i = v(\mathcal{N})$. Any unilateral deviation $\sigma'_i$ that attempts to increase $\mu_i$ must either (1) modify the committed output, which is detected with probability $1 - \text{negl}(\lambda)$ by G2, or (2) manipulate the Shapley computation, which is verified by the ZKP (G3). In both cases, the deviation triggers verification failure, resulting in zero payout. Thus, $\mathbb{E}[\text{reward}_i(\sigma'_i)] \leq \text{negl}(\lambda) \cdot v(\mathcal{N}) < \mathbb{E}[\phi_i]$.
    \end{proof}
  }

\subsubsection{Goal G2: Cryptographic Commitment for Output Authenticity}
\label{subsubsec:goal-commitment}

Challenge C2 demands a verifiable link between off-chain execution artifacts and on-chain settlement.

\textbf{Guarantee G2 (Commitment Binding).}\label{property:commitment-binding}
Once an agent output $o_i$ is committed at time $t$, any modification $o'_i \neq o_i$ is detected and rejected with probability at least $1 - \text{negl}(\lambda)$. For any coalition $S \subseteq \mathcal{N}$ with the associated output set $O_S = \{o_i : a_i \in S\}$ (the collection of the outputs from the agents in $S$) and the corresponding output hash set $\mathcal{H}_{O_S} = \{H(o_i) : o_i \in O_S\}$ (the set of the cryptographic hashes of these outputs),
\begin{equation}
  \Pr[\text{Accept}(O'_S) \mid \exists o'_i \in O'_S : H(o'_i) \notin \mathcal{H}_{O_S}] \leq \text{negl}(\lambda).
\end{equation}
where $O'_S$ is the submitted output set for coalition $S$, and $\text{Accept}(\cdot)$ is the verification predicate that returns true if the submitted outputs are consistent with the on-chain commitment \cite{olimid2025commitment,chainlink2023commit}.

This guarantee is realized through a dual-commitment scheme that provides a trustless foundation for off-chain computation. First, agent outputs ($o_i$) are deterministically captured and hashed; the resulting hash set is stored on IPFS to generate a content identifier (CID), which is then anchored on-chain \cite{benet2014ipfs,trautwein2022design}. Second, the evaluated coalition value ($v(S)$) is also hashed, and its hash ($H(v(S))$) is committed directly to the blockchain. This layered cryptographic binding ensures that any tampering with either the outputs or their evaluated value is detected when the Coordinator verifies the data against the immutable on-chain anchors (as detailed in Algorithm~\ref{alg:hybrid-workflow}). Security relies on the collision resistance of the hash function $H$ and the immutability of the blockchain.

  {
    \begin{theorem}[Commitment Binding]
      For any adversary $\mathcal{A}$ running in polynomial time, the probability of producing a valid output set $O'_S$ inconsistent with the on-chain commitment is negligible: $\Pr[\text{Accept}(O'_S) \land \exists o'_i : H(o'_i) \notin \mathcal{H}_{O_S}] \leq \text{negl}(\lambda)$.
    \end{theorem}

    \begin{proof}[Proof Sketch]
      Assume $\mathcal{A}$ produces $o'_i \neq o_i$ such that $H(o'_i) = H(o_i)$. This implies a hash collision, which contradicts the collision resistance of $H$ under standard assumptions. Since the hash set $\mathcal{H}_{O_S}$ is anchored on the immutable blockchain, any modification to $O_S$ after commitment requires finding a collision, occurring with probability at most $\text{negl}(\lambda)$.
    \end{proof}
  }

\subsubsection{Goal G3: Zero-Knowledge Proofs for Scalable Verification}
\label{subsubsec:goal-zkp}

Challenge C3 requires verifiable computation without transferring exponential costs to the blockchain.

\textbf{Guarantee G3 (Computational Scalability).}\label{property:computational-scalability}
Verification cost remains $O(1)$ as agent count $n$ grows, even when contribution measurement requires $O(2^n)$ coalition evaluations. This decouples on-chain verification from off-chain computational complexity, enabling practical deployment at scale.

The guarantee is realized through a hybrid proof pipeline: the off-chain prover executes the full Shapley computation off-chain, produces a STARK proof, and recursively compresses it into a constant-size Groth16 proof for on-chain verification (Section~\ref{subsec:zkp-integration}) \cite{groth2016size}. Consequently, the blockchain only performs a fixed-cost pairing check while inheriting soundness from the underlying proof systems.

  {
    \begin{theorem}[Verification Soundness]
      For any computationally bounded adversary, the probability of the on-chain verifier accepting an incorrect Shapley allocation is negligible: $\Pr[f_{\text{verify}}(\pi, pub) = 1 \land \exists i : \mu_i \neq \phi_i] \leq \text{negl}(\lambda)$.
    \end{theorem}

    \begin{proof}[Proof Sketch]
      The ZKP circuit enforces constraint (iii): $\forall i: \mu_i = \phi_i(\{v(S)\})$. By the knowledge soundness of the Groth16 proof system, any accepting proof $\pi$ implies the existence of a valid witness satisfying all constraints. If $\mu_i \neq \phi_i$, constraint (iii) is violated, and no valid proof exists. The probability of forging a proof without a valid witness is bounded by the soundness error of Groth16, which is $\text{negl}(\lambda)$ under the $q$-PKE assumption \cite{groth2016size}.
    \end{proof}
  }

Having established the system model, identified core challenges, and formalized design goals with security guarantees, we now present the concrete architectural mechanisms that realize these properties in Section~\ref{sec:framework}.

\begin{figure*}
  \centering
  \includegraphics[width=1.0\textwidth]{DAO-Agents-Framework.png}
  \caption{The DAO-Agent hybrid architecture, integrating off-chain collaborative execution with on-chain cryptographic verification. The framework bridges the trust gap through a four-phase workflow: 
  (1) \textbf{Off-Chain Execution and Multi-Tier Commitment}, where agent outputs $O_S$ and coalition values $v(S)$ are hashed ($\mathcal{H}_{O_S}, \mathcal{H}_{v(S)}$) and anchored on-chain via IPFS CIDs; 
  (2) \textbf{Coordinator Integrity Verification}, where the Coordinator trustlessly validates hash integrity and evaluates axiomatically fair rewards using the Shapley value $\phi_i$ while enforcing the Efficiency Axiom ($\sum \mu_i = v(\mathcal{N})$); 
  (3) \textbf{Recursive Proof Composition}, involving the generation of a STARK proof for the $O(2^n)$ Shapley computation, subsequently compressed into a constant-size Groth16 proof for blockchain compatibility; and 
  (4) \textbf{On-Chain Settlement}, where the verification contract performs a single $O(1)$ pairing check to trigger automated reward distribution.}
  \Description{System architecture diagram showing the DAO-Agent hybrid framework with four phases: execution with multi-tier hash commitments, integrity verification with Shapley allocation, STARK-to-Groth16 proof composition, and constant-cost on-chain verification.}
  \label{fig:framework-architecture}
\end{figure*}

\section{DAO-Agent Framework Design}
\label{sec:framework}

To address the challenges of fair contribution measurement (C1), output authenticity (C2), and scalable verification (C3), we propose the framework of DAO-Agent. As illustrated in Figure~\ref{fig:framework-architecture}, our design is a hybrid on-chain/off-chain architecture that synergizes three core technical components: (1) an on-chain DAO for governance and settlement, (2) an off-chain workflow for agent execution and contribution measurement, and (3) a zero-knowledge proof (ZKP) mechanism cryptographically link the two. This separation of concerns allows computationally intensive tasks to execute off-chain while leveraging the blockchain for what it does best: providing a trustless source of truth for verification and settlement \cite{yang2024swarm,qi2025towards}.

The on-chain component serves three critical functions that distinguish DAO-Agent from traditional off-chain computation approaches: (1) \textbf{Governance Anchor}: immutable task lifecycle management and agent registration via smart contract state machine, ensuring transparent coordination with cryptographic certainty; (2) \textbf{Verification Layer}: constant-cost ZKP verification (approximately 27k gas) that remains independent of agent count $n$, enabling scalable trust without prohibitive on-chain expenses; (3) \textbf{Settlement Engine}: cryptographically certain incentive distribution with on-chain transparency, eliminating the need for trusted third parties while preserving execution privacy. This hybrid model represents a paradigm shift from direct on-chain computation—where Shapley value calculation would require $O(2^n)$ gas—to verification-centric architecture that inherits blockchain's trust properties while avoiding its computational limitations.

\subsection{Workflow-Based Commitments and Hybrid Trust Pipeline}
\label{subsec:workflow-and-pipeline}

The DAO-Agent framework implements a comprehensive hybrid workflow that orchestrates the entire lifecycle of decentralized multi-agent collaboration through cryptographically secure coordination between off-chain execution and on-chain verification. Algorithm~\ref{alg:hybrid-workflow} presents the complete four-phase protocol that systematically addresses the fundamental challenges of fair contribution measurement, output authenticity, and scalable verification identified in Section~\ref{subsec:challenges}.

\begin{algorithm}[t]
  \caption{DAO-Agent Hybrid Workflow}
  \label{alg:hybrid-workflow}
  \KwIn{Task $T$, Agent Set $\mathcal{N} = \{a_1,\ldots,a_n\}$, Coordinator $\mathcal{C}$}

  \tcp{Phase 1: Subset Execution and Hash Commitment}
  \ForEach{coalition $S \subseteq \mathcal{N}$}{
    Agents in $S$ execute $T$ in workflow environment\;
    $O_S \gets \{o_i : a_i \in S\}$ \tcp{Coalition $S$ output set}
    $\mathcal{H}_{O_S} \gets \{H(o_i) : o_i \in O_S\}$ \tcp{Hash set of outputs}
    $\text{CID}_{O_S} \gets f_{\text{IPFS}}(\mathcal{H}_{O_S})$
    $\mathcal{E}_{\text{commit}}(\text{CID}_{O_S}) \to (t_{\text{block}}, \text{txHash})$
    $v(S) \gets f_{\text{eval}}(O_S)$ \tcp{Coalition value evaluation}
    $\mathcal{H}_{v(S)} \gets H(v(S))$ \tcp{Hash of coalition value}
    $\mathcal{E}_{\text{commitValue}}(\mathcal{H}_{v(S)}) \to (t_{\text{block}}', \text{txHash}')$
  }

  \BlankLine
  \tcp{Phase 2: Coordinator Verification and Shapley Computation}
  \ForEach{received $O_S$}{
    $\text{CID}_{O_S} \gets \mathcal{E}_{\text{getCID}}(S)$
    \textbf{assert} $\{H(o_i) : o_i \in O_S\} = \text{IPFS.get}(\text{CID}_{O_S})$\;
  }
  \ForEach{received $v(S)$}{
    $\mathcal{H}_{v(S)} \gets \mathcal{E}_{\text{getValueHash}}(S)$
    \textbf{assert} $H(v(S)) = \mathcal{H}_{v(S)}$
  }
  $\{\mu_i : a_i \in \mathcal{N}\} \gets \frac{1}{n} \sum_{S \subseteq \mathcal{N}\setminus\{a_i\}} \binom{n-1}{|S|}^{-1} \bigl(v(S \cup \{a_i\}) - v(S)\bigr)$ \tcp{Shapley allocation}

  \BlankLine
  \tcp{Phase 3: Zero-Knowledge Proof Generation}
  $\pi \gets f_{\text{proveZK}}(\{\mu_i : a_i \in \mathcal{N}\}, \{\mathcal{H}_{O_S}, \mathcal{H}_{v(S)}, O_S, v(S)\})$\;

  \BlankLine
  \tcp{Phase 4: On-Chain Verification and Settlement}
  $pub\_inputs \gets (\{\mu_i : a_i \in \mathcal{N}\}, v(\mathcal{N}), \{\mathcal{H}_{O_S} : S \subseteq \mathcal{N}\}, \{\mathcal{H}_{v(S)} : S \subseteq \mathcal{N}\})$\;
  \If{$f_{\text{verify}}(\pi, pub\_inputs, \mathcal{E}) = \text{true}$}{
    $f_{\text{allocate}}(\{\mu_i : a_i \in \mathcal{N}\}) \to \mathcal{E}$\;
    \Return $\{\mu_i : a_i \in \mathcal{N}\}$\;
  }
  \Return \textbf{abort}\;
\end{algorithm}

\textbf{Hybrid Protocol Overview.} Algorithm~\ref{alg:hybrid-workflow} orchestrates the complete workflow through four distinct phases that systematically address the challenges of fair contribution measurement (C1), output authenticity (C2), and scalable verification (C3). The protocol maintains a clear separation of concerns: computationally intensive tasks execute off-chain while the blockchain provides trustless verification and settlement with cryptographic certainty.

\textbf{Cryptographic Infrastructure.} The algorithmic realization of our hybrid architecture relies on a carefully orchestrated set of cryptographic primitives and smart contract interfaces that bridge the gap between off-chain computation and on-chain verification. The IPFS decentralized storage system serves as our content-addressed persistence layer, where $f_{\text{IPFS}}(\cdot)$ atomically stores hash sets and returns content identifiers (CIDs) that serve as tamper-proof references to agent outputs. These CIDs are subsequently anchored on-chain through the smart contract's commitment interface $\mathcal{E}_{\text{commit}}(\cdot)$, which generates immutable timestamps $t_{\text{block}}$ and transaction hashes $\text{txHash}$ that provide cryptographic binding for all off-chain artifacts. The dual-retrieval mechanism $\mathcal{E}_{\text{getCID}}(S)$ and $\mathcal{E}_{\text{getValueHash}}(S)$ enables the Coordinator to trustlessly reconstruct the complete commitment state, while $\text{IPFS.get}(\cdot)$ resolves CIDs to their corresponding hash sets for integrity verification. Zero-knowledge proof generation through $f_{\text{proveZK}}(\cdot)$ transforms the exponentially complex Shapley computation into succinct cryptographic arguments, and the verification function $f_{\text{verify}}(\cdot)$ provides constant-cost on-chain validation that remains independent of the underlying computational complexity. Finally, the incentive distribution mechanism $f_{\text{allocate}}(\cdot)$ atomically transfers rewards according to the proven-fair allocation, completing the trustless coordination cycle while maintaining execution privacy and computational efficiency.

\textbf{Phase 1: Subset Execution and Hash Commitment.} Every possible agent coalition $S \subseteq \mathcal{N}$ executes the collaborative task off-chain, producing outputs that are immediately cryptographically secured. This phase implements the Commitment Binding guarantee (G2) through a dual-commitment scheme that prevents output manipulation without relying on agent self-reporting.

\textbf{Phase 2: Coordinator Verification and Shapley Computation.} The Coordinator acts as a trustless computation node that verifies the integrity of all committed data and computes axiomatically fair Shapley value allocations. This phase realizes the Incentive Compatibility guarantee (G1) by ensuring that contribution measurements are both cryptographically verifiable and game-theoretically sound.

\textbf{Phase 3: Zero-Knowledge Proof Generation.} Zero-knowledge proofs are generated to attest to the correctness of the entire off-chain computation without revealing sensitive execution details. This phase provides the Computational Scalability guarantee (G3) by converting exponential off-chain computation into constant-size cryptographic proofs.

\textbf{Phase 4: On-Chain Verification and Settlement.} The smart contract performs efficient on-chain verification and atomically distributes incentives according to the proven-fair allocation. This final phase ensures that all previous computations are cryptographically validated before any value transfer occurs.

The framework's operation is orchestrated by the smart contract $\mathcal{E}$, which implements a hybrid trust model combining off-chain computational efficiency with on-chain verification integrity. On-chain verification consumes approximately 27k gas regardless of agent count $n$, achieving constant $O(1)$ complexity through succinct non-interactive argument verification. This represents a paradigm shift from traditional on-chain computation—where exact Shapley value calculation would require exponential $O(2^n)$ gas consumption—to a verification-centric approach that inherits blockchain's trust properties while avoiding its computational limitations. Our hybrid architecture reduces on-chain verification complexity from exponential to constant, enabling practical deployment for large-scale multi-agent coordination while maintaining cryptographic certainty.

\textbf{From Protocol to Mechanisms.} While Algorithm~\ref{alg:hybrid-workflow} presents the complete workflow coordination, three fundamental technical mechanisms enable this hybrid architecture to achieve its security and efficiency guarantees. First, the protocol requires \textbf{fair contribution measurement} that can withstand strategic manipulation by rational agents—this necessitates a rigorous game-theoretic foundation that we develop in Section~\ref{subsec:shapley}. Second, the Coordinator must generate \textbf{cryptographic proofs} that convincingly demonstrate computation integrity without revealing sensitive execution data—this demands sophisticated zero-knowledge proof integration detailed in Section~\ref{subsec:zkp-integration}. Third, the entire system must maintain \textbf{computational scalability} as the number of agents grows, requiring careful architectural decisions that separate expensive off-chain computation from lightweight on-chain verification.

The following sections dissect these three critical mechanisms. Section~\ref{subsec:shapley} establishes the game-theoretic foundation by demonstrating how Shapley values provide axiomatically fair contribution measurement that makes honest participation a dominant strategy for agents. Section~\ref{subsec:zkp-integration} then shows how zero-knowledge proofs cryptographically link the off-chain Shapley computation to on-chain verification, ensuring that fairness guarantees are preserved while maintaining execution privacy. Together, these mechanisms transform the high-level workflow presented above into a concrete, implementable system that resolves the tension between decentralization, privacy, and computational efficiency in multi-agent coordination.

\subsection{Shapley-Based Contribution Measurement}
\label{subsec:shapley}

To achieve fair contribution measurement (C1) and realize the Incentive Compatibility guarantee (G1), we ground our reward mechanism in the Shapley value from cooperative game theory. The Shapley value, $\phi_i(v)$, provides an axiomatically fair allocation by quantifying each agent $a_i$'s contribution as its average marginal value across all possible coalition permutations.

Formally, for each agent $a_i \in \mathcal{N}$, the Shapley value is:
\begin{equation}
  \label{eq:shapley}
  \phi_i(\{v(S) : S \subseteq \mathcal{N}\})
  = \frac{1}{n}\!
  \sum_{S \subseteq \mathcal{N}\setminus\{a_i\}}
  \binom{n-1}{|S|}^{-1}
  \bigl(v(S \cup \{a_i\}) - v(S)\bigr)\end{equation}
The weight term ensures that all agent joining orders are equally likely, eliminating order bias. This method yields a unique allocation satisfying four axioms: \textbf{Efficiency} ($\sum_{i \in \mathcal{N}} \phi_i = v(\mathcal{N})$), \textbf{Symmetry}, \textbf{Dummy}, and \textbf{Additivity}. The uniqueness of this allocation provides an indisputable foundation for fair rewards, making honest participation a dominant strategy when combined with the cryptographic guarantees that follow. The efficiency axiom is particularly central to our design, serving as: (1) a theoretical guarantee of complete value distribution, (2) a computational constraint verified within the ZKP, and (3) a final on-chain check using public inputs. To formalize this verification, we distinguish between the true Shapley value, defined by the function $\phi_i$, and the concrete allocation vector computed by the Coordinator, which we denote as $\{\mu_i\}$. The core of the ZKP is to prove that the submitted $\{\mu_i\}$ correctly corresponds to the output of the $\phi_i$ function for all agents (i.e., that $\mu_i = \phi_i$).

Within our hybrid architecture, Shapley computation is performed off-chain by the Coordinator, leveraging the computational efficiency of traditional computing environments while maintaining verifiability through on-chain commitments. The on-chain smart contract $\mathcal{E}$ does not perform the expensive $O(2^n)$ coalition evaluations directly; instead, it verifies that the off-chain computation was performed correctly on the committed data through zero-knowledge proof verification. This separation enables the system to achieve both computational scalability and cryptographic verifiability without compromising either property.

\begin{algorithm}[htbp]
  \caption{ZKP Construction: Witness Generation and Constraint Enforcement}
  \label{alg:zkp-generation}
  \KwIn{Coalition Outputs $\{O_S\}$, Coalition Values $\{v(S)\}$, Shapley Allocation $\{\mu_i\}$, Commitments $\{\mathcal{H}_{O_S}, \mathcal{H}_{v(S)}\}$}
  \KwOut{Proof $\pi$ and public inputs}

  \tcp{\textbf{Phase I: Off-Chain Witness Computation}}
  \tcp{Generates the execution trace $\mathcal{W}$ for the arithmetic circuit}
  $\mathcal{W} \leftarrow \text{InitTrace}()$\;
  \ForEach{$S \subseteq \mathcal{N}$}{
    $\mathcal{W}.\text{AppendInput}(O_S, v(S))$\;
  }
  \ForEach{agent $a_i \in \mathcal{N}$}{
    $\phi_i \leftarrow \text{ComputeShapley}(i, \{v(S)\})$ \tcp{Execute Eq. 3}
    $\mathcal{W}.\text{RecordComputation}(i, \phi_i)$\;
  }

  \BlankLine
  \tcp{\textbf{Phase II: In-Circuit Constraint Enforcement}}
  \tcp{Arithmetic constraints verified within the ZK circuit}
  \SetKwBlock{Circuit}{Circuit Logic:}{end}
  \Circuit{
    \tcp{C1: Input Consistency Check}
    \ForEach{$S \subseteq \mathcal{N}$}{
      \textbf{constrain} $\text{PoseidonHash}(\mathcal{W}.O_S) \equiv public\_input.\mathcal{H}_{O_S}$\;
      \textbf{constrain} $\text{PoseidonHash}(\mathcal{W}.v(S)) \equiv public\_input.\mathcal{H}_{v(S)}$\;
    }
    
    \tcp{C2: Shapley Arithmetization}
    \ForEach{agent $a_i \in \mathcal{N}$}{
       \tcp{Polynomial constraints for Eq. 3}
       $sum \leftarrow 0$\;
       \ForEach{$S \subseteq \mathcal{N}\setminus\{a_i\}$}{
         $coeff \leftarrow \text{Lookup}(\binom{n-1}{|S|}^{-1})$\;
         $marginal \leftarrow \mathcal{W}.v(S \cup \{a_i\}) - \mathcal{W}.v(S)$\;
         $sum \leftarrow sum + (coeff \times marginal)$\;
       }
       \textbf{constrain} $\mathcal{W}.\mu_i \equiv sum \times n^{-1}$\;
    }

    \tcp{C3: Efficiency Axiom Check}
    \textbf{constrain} $\sum \mathcal{W}.\mu_i \equiv \mathcal{W}.v(\mathcal{N})$\;
  }

  \BlankLine
  \tcp{\textbf{Phase III: Recursive Proof Generation}}
  $\pi \leftarrow \text{STARK.Prove}(Circuit, \mathcal{W}, public\_inputs)$\;
  \Return $(\pi, public\_inputs)$\;
\end{algorithm}

\subsection{Zero-Knowledge Proof Integration}
\label{subsec:zkp-integration}

To address scalable, privacy-preserving verification (C3) and provide the Computational Scalability guarantee (G3), DAO-Agent integrates zero-knowledge proofs. This allows the off-chain Coordinator to prove the correctness of the computationally expensive Shapley value calculation without revealing any private data (e.g., agent strategies embedded in outputs) and without imposing exponential costs on the blockchain. The on-chain verification component consumes approximately 27k gas regardless of agent count, achieving the $O(1)$ complexity required for practical deployment at scale.

\subsubsection{Proof of Correct Computation}
\label{subsubsec:proof-protocol}

The Coordinator uses a proving function of ZKP to generate a proof $\pi$ attesting to the integrity of the entire off-chain collaboration process. The proving function of ZKP takes public input and private witness as the inputs and produces a proof for the computation correctness. The verification function of ZKP requires only the proof and public inputs to confirm the correctness of the computation, achieving this with significantly lower complexity compared to re-executing the entire computation process.

Specifically, the ZKP protocol is defined over a relation $\mathcal{R}$ comprising a private witness $\Omega_{\text{private}}$ and public input $\Omega_{\text{public}}$:
\begin{align}
  \left\{\begin{aligned}
           \Omega_{\text{private}} & = (\{O_S\}_{S \subseteq \mathcal{N}}, \{v(S)\}_{S \subseteq \mathcal{N}}),                                                       \\
           \Omega_{\text{public}}  & = (\{\mu_i\}, v(\mathcal{N}), \{\mathcal{H}_{O_S}\}, \{\mathcal{H}_{v(S)}\})
         \end{aligned}\right.
\end{align}
Here, $\Omega_{\text{private}}$ encapsulates the complete execution trace, including all coalition outputs $O_S$ and their corresponding evaluations $v(S)$. The public input $\Omega_{\text{public}}$ serves as the on-chain anchor, containing the proposed Shapley allocation $\{\mu_i\}$, the grand coalition value $v(\mathcal{N})$, and the set of committed cryptographic hashes. The validity of the proof $\pi$ is contingent upon satisfying the following arithmetic constraints:
\begin{align}
  \left\{\begin{aligned}
           \text{(i)} ~  & \forall S \subseteq \mathcal{N}: \{H(o_i) : o_i \in O_S\} = \mathcal{H}_{O_S}, \\
           \text{(ii)} ~ & \forall S \subseteq \mathcal{N}: H(v(S)) = \mathcal{H}_{v(S)},                 \\
           \text{(iii)}~ & \forall i: \mu_i= \phi_i(\{v(S) : S \subseteq \mathcal{N}\}),                  \\
           \text{(iv)} ~ & \sum_{i \in \mathcal{N}} \mu_i = v(\mathcal{N})
         \end{aligned}\right.
\end{align}
These constraints enforce the system's core integrity properties:
\begin{enumerate}
  \item[(i)] \textbf{Input Consistency:} guarantees that the private witness values correspond exactly to the immutable on-chain commitments, preventing retroactive data manipulation.
  \item[(ii)] \textbf{Value Authenticity:} ensures the coalition evaluations used in the circuit match the committed values.
  \item[(iii)] \textbf{Computational Correctness (Shapley Arithmetization):} validates that the allocation $\{\mu_i\}$ is derived via the exact Shapley formula (Eq.~\eqref{eq:shapley}). We arithmetize the iterative summation and combinatorial weighting logic into polynomial constraints (see Algorithm~\ref{alg:zkp-generation}, Lines 13-20), thereby proving that the $O(2^n)$ factorials and marginal contributions were computed correctly.
  \item[(iv)] \textbf{Conservation of Value:} enforces the Efficiency Axiom, ensuring the total distributed reward strictly equals the generated value.
\end{enumerate}

These cryptographic assertions form the basis of our "Code is Law" implementation, distinguishing DAO-Agent from systems relying on optimistic assumptions or centralized verifiers. Algorithm~\ref{alg:zkp-generation} delineates the formal separation between off-chain trace generation and in-circuit constraint enforcement.

  \subsubsection{Recursive Proof Composition and On-Chain Verification}
\label{subsubsec:proof-composition}

To reconcile the computational complexity of Shapley value derivation with the strict gas limits of blockchain execution, we employ a recursive proof composition strategy. The pipeline initiates with the generation of STARK proofs (Scalable Transparent Arguments of Knowledge), which provide the necessary computational capacity for checking $O(2^n)$ coalition values without trusted setup. However, the resulting proofs are typically large (tens of kilobytes), rendering direct on-chain verification economically infeasible.

To address this, we utilize a proof-carrying data (PCD) approach where the STARK proof $\pi_{\text{STARK}}$ is recursively verified within a secondary circuit. This circuit generates a succinct Groth16 proof $\pi_{\text{SNARK}}$, compressing the verification complexity to constant size. This STARK-to-SNARK recursion leverages the transparency and scalability of STARKs for the heavy arithmetic lifting (Shapley computation) while utilizing the compactness of pairing-based SNARKs for final on-chain settlement. Consequently, the smart contract performs a single $O(1)$ pairing check ($\approx$ 27k gas), decoupling the on-chain verification cost from the exponential off-chain workload.

\section{Experimental Evaluation}
\label{sec:experiments}

\subsection{Experimental Setup}
\label{subsec:exp-setup}

We implement a multi-agent cryptocurrency trading system with a fixed architecture (2 data analysis agents, 1 decision-making agent, and a variable number of market perspective agents, where $x \in \{1, 3, 5, 7\}$) deployed on an Ethereum testnet. For each coalition $S \subseteq \mathcal{N}$, we quantify its value $v(S)$ using Sharpe ratio-based trading performance metrics over ETH/USDT historical data from 2022-2024. The experiments evaluate agent sets of size $n \in \{4, 6, 8, 10\}$, requiring $2^n$ coalition evaluations each.

Our performance assessment is structured into two key areas: the off-chain computational overhead required for proof generation and the on-chain resource consumption for verification. We evaluate DAO-Agent against an \textit{On-chain Shapley} baseline, which performs all computations directly in a smart contract. The coalition value function $v(S)$ is computed using Sharpe ratio-based trading performance metrics, validated against historical ground-truth data. The results validate our hybrid model's ability to achieve constant-cost on-chain scalability while managing exponential complexity off-chain.

\begin{figure}[t]
  \centering
  \bfseries{
    \includegraphics[width=0.48\textwidth]{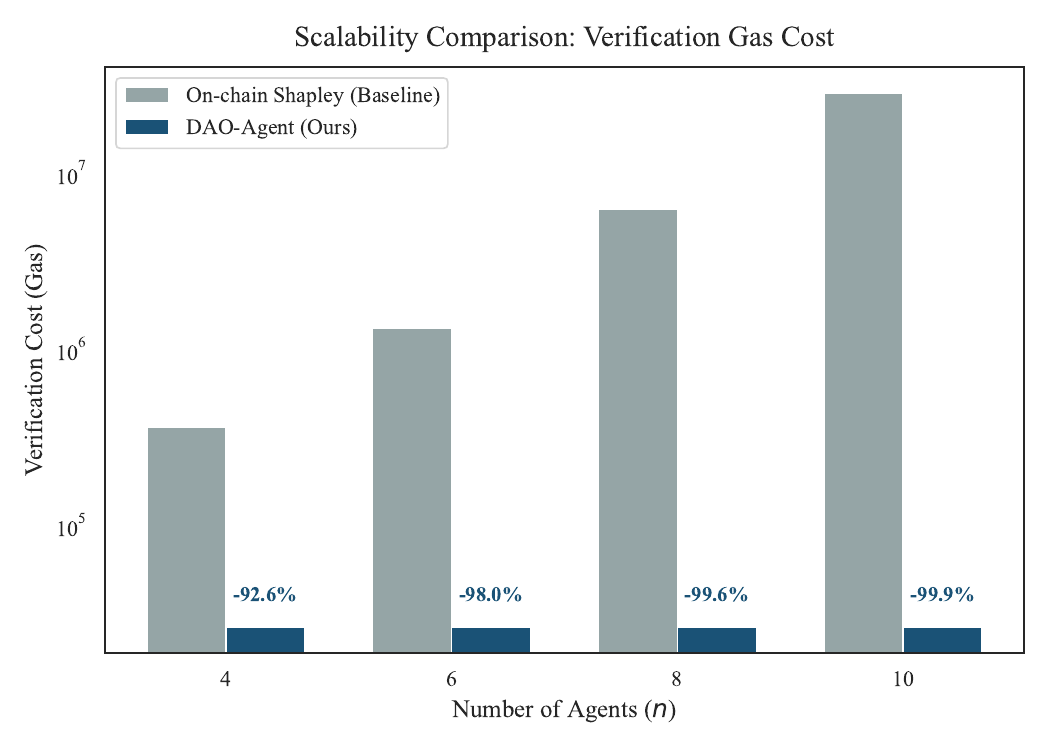}
    \caption{Scalability Comparison: Verification Gas Cost (Log Scale). The bar chart demonstrates a massive cost reduction of up to 99.9\% for larger agent coalition ($n=10$), contrasting the exponential growth of the baseline with the constant cost of our hybrid approach.}
    \Description{Bar chart comparing gas costs between on-chain and hybrid architecture across different agent counts, showing exponential growth for on-chain versus constant cost for hybrid approach.}
    \label{fig:gas-cost-comparison}
  }
\end{figure}

\begin{figure}[t]
  \centering
  \bfseries{
    \includegraphics[width=0.48\textwidth]{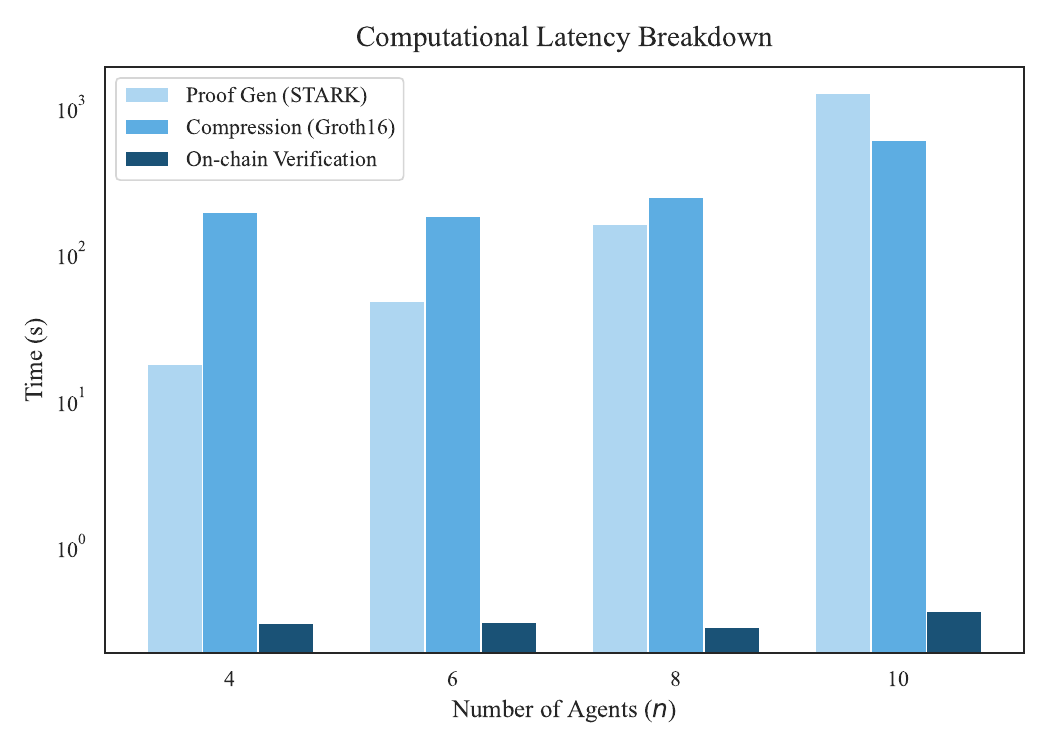}
    \caption{Computational Latency Breakdown (Log Scale). The grouped bar chart reveals the trade-off: while off-chain proof generation (STARK) scales exponentially, the on-chain verification time remains negligible and constant ($<0.4s$).}
    \Description{Grouped bar chart showing ZKP generation time breakdown: exponential growth for STARK proof generation, near-constant time for Groth16 conversion, and constant verification time across agent counts.}
    \label{fig:zkp-time-breakdown}
  }
\end{figure}

\subsection{Off-Chain Computational Performance}
\label{subsec:exp-offchain}

The primary computational workload of DAO-Agent resides in the off-chain proof generation pipeline (Phase 3). As detailed in Table~\ref{tab:zkp-performance} and Figure~\ref{fig:zkp-time-breakdown}, the STARK proof generation time is the main bottleneck, scaling exponentially with the number of agents. This duration increases from 21.22 seconds for 4 agents to 425.60 seconds for 10 agents, reflecting the $O(2^n)$ complexity of the underlying Shapley value computation.

In contrast, the subsequent proof conversion to Groth16 remains highly efficient and does not exhibit exponential growth. Furthermore, the final proof size increases only marginally, from 1,417 bytes to 1,802 bytes, ensuring that the cryptographic artifacts remain compact and manageable. These results pinpoint the STARK generation as the key scalability constraint of the off-chain process.

\begin{table}[h]
  \centering
  \caption{ZKP Performance: Proof Generation and Verification Metrics}
  \label{tab:zkp-performance}
  \setlength{\tabcolsep}{3pt}
  \begin{tabular}{@{}lcccc@{}}
    \toprule
    \textbf{Agents} & \textbf{STARK (s)} & \textbf{Groth16 (s)} & \textbf{Size (B)} & \textbf{Verify (s)} \\
    \midrule
    4               & 21.22              & 233.41               & 1,417             & 0.358               \\
    6               & 56.46              & 217.19               & 1,545             & 0.368               \\
    8               & 201.36             & 310.27               & 1,673             & 0.354               \\
    10              & 425.60             & 298.15               & 1,802             & 0.362               \\
    \bottomrule
  \end{tabular}
\end{table}

\subsection{On-Chain Cost and Scalability}
\label{subsec:exp-onchain}

The central advantage of the DAO-Agent framework is its ability to achieve constant-cost on-chain scalability. Figure~\ref{fig:gas-cost-comparison} illustrates the dramatic difference between our hybrid approach and the on-chain baseline. While the baseline's gas cost explodes exponentially, DAO-Agent's verification cost remains flat at approximately 27k gas, regardless of the number of agents.

As shown in Table~\ref{tab:gas-comparison}, this translates to a gas reduction of 99.9\% for a 10-agent system (27k vs. 28.6M gas). This efficiency transforms an intractable $O(2^n)$ on-chain problem into a lightweight $O(1)$ verification. The on-chain verification itself is consistently fast, taking approximately 0.36 seconds (see Table~\ref{tab:zkp-performance}). This low, predictable gas cost, comparable to a standard ERC-20 transfer, confirms that DAO-Agent is economically viable for practical deployment at scale.

\begin{figure}[t]
  \centering
  \bfseries{
    \includegraphics[width=0.48\textwidth]{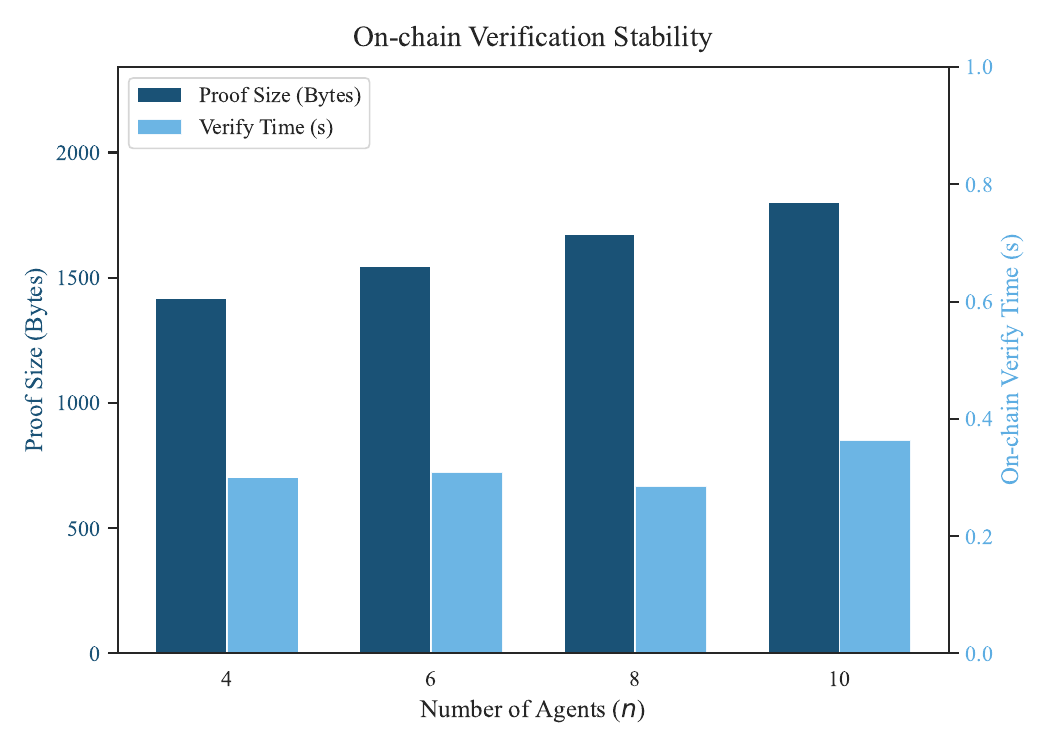}
    \caption{On-chain Verification Stability. The dual-axis bar chart illustrates that both Proof Size (Left Axis) and Verification Time (Right Axis) remain stable and low despite the increasing complexity of the multi-agent task.}
    \Description{Dual-axis bar chart showing constant proof size and verification time across increasing agent counts.}
    \label{fig:verification-stability}
  }
\end{figure}

\begin{table}[h]
  \centering
  \caption{Gas Cost Comparison: On-chain vs. DAO-Agent Verification}
  \label{tab:gas-comparison}
  \begin{tabular}{@{}lccc@{}}
    \toprule
    \textbf{Agents} & \textbf{On-chain Gas} & \textbf{DAO-Agent Gas} & \textbf{Reduction} \\
    \midrule
    4               & 367K                  & 27K                    & 92.6\%             \\
    6               & 1.33M                 & 27K                    & 98.0\%             \\
    8               & 6.35M                 & 27K                    & 99.6\%             \\
    10              & 28.6M                 & 27K                    & 99.9\%             \\
    \bottomrule
  \end{tabular}
\end{table}

\section{Conclusion}
\label{sec:conclusion}

This paper addressed the core challenge of achieving fair, scalable, and privacy-preserving coordination in decentralized multi-agent systems, which is hindered by prohibitive on-chain computation costs and the lack of verifiable contribution measurement. We introduced DAO-Agent, a novel hybrid on-chain/off-chain framework that leverages the blockchain as a trust anchor while offloading computationally intensive tasks. Our approach cryptographically links Shapley value-based contribution measurements to a constant-cost on-chain verification mechanism using zero-knowledge proofs.

Our experimental evaluation demonstrates the practicality and efficiency of DAO-Agent. By transforming the exponential $O(2^n)$ on-chain complexity into a constant $O(1)$ verification process, our method reduces gas costs by up to 99.9\% for 10 agents compared to a purely on-chain implementation. This breakthrough makes fair, trustless multi-agent coordination economically viable in decentralized environments.

  While our experiments validate efficiency for up to 10 agents, the architecture is designed for extensibility. The off-chain proof generation complexity ($O(2^n)$ for exact Shapley) can be addressed through approximation techniques such as Monte Carlo Shapley estimation, enabling larger-scale deployments with minimal precision trade-offs. Furthermore, the system's security model, which currently assumes coordinator liveness, could be strengthened by introducing redundant coordinators or decentralized challenge-response mechanisms. The security framework relies on standard cryptographic assumptions and is adaptable to post-quantum hash functions for future-proofing.

In summary, DAO-Agent provides a robust and efficient blueprint for building trustless, incentive-compatible multi-agent systems. By offering substantial cost savings for frequent coordination tasks, it paves the way for more sophisticated and secure decentralized collaborations. Future work could further enhance performance by exploring layer-2 deployment and proof batching techniques.

\bibliographystyle{ACM-Reference-Format}
\bibliography{reference}

\end{document}